\documentclass[12pt]{iopart}

\usepackage{iopams}
\usepackage{graphicx}
\usepackage{float}
\begin{document}

\title[]{Imaging magnetoelectric subbands in ballistic constrictions}

\author{A.~A.~Kozikov$^1$, D. Weinmann$^2$, C.~R\"{o}ssler$^1$, T.~Ihn$^1$, K.~Ensslin$^1$, C.~Reichl$^1$ and W.~Wegscheider$^1$}

\address{$^1$ Solid State Physics Laboratory, ETH Z\"{u}rich, CH-8093 Z\"{u}rich, Switzerland}
\address{$^2$ Institut de Physique et Chimie des Mat\'{e}riaux de Strasbourg,
Universit\'{e} de Strasbourg, CNRS UMR 7504, 23 rue du Loess, F-67034 Strasbourg, France}
\ead{akozikov@phys.ethz.ch}
\begin{abstract}

We perform scanning gate experiments on ballistic constrictions in the presence of small perpendicular magnetic fields. The constrictions form the entrance and exit of a circular gate-defined ballistic stadium. Close to constrictions we observe sets of regular fringes creating a checker board pattern. Inside the stadium conductance fluctuations governed by chaotic dynamics of electrons are visible. The checker board pattern allows us to determine the number of transmitted modes in the constrictions forming between the tip-induced potential and gate-defined geometry. Spatial investigation of the fringe pattern in a perpendicular magnetic field shows a transition from electrostatic to magnetic depopulation of magnetoelectric subbands. Classical and quantum simulations agree well with different aspects of our observations.

\end{abstract}


\maketitle

\section{Introduction}

Quantum point contacts are the basic building blocks of many mesoscopic circuits. Quantized conductance requires ballistic transport with an elastic mean free path exceeding the dimensions of the quantum point contact. In a series arrangement of two quantum point contacts \cite{QPCseries} the total conductance depends on how electron waves couple into and out of the region between the quantum point contacts. If this region is a cavity, then interference of electron waves can also play an important role.
In mesoscopic devices the gate arrangement and therefore the dominant features of the potential landscape are usually fixed. A moving gate, as it is provided by the metallic tip of an atomic force microscope (AFM), allows us to vary the potential landscape in real space and enables the investigation of ballistic and coherent transport properties of networks of quantum point contacts (QPC). Not only the tip-sample bias voltage, but also the tip-surface distance and the in-plane tip position can be changed in order to control the strength and gradient of the tip-induced potential and thereby influence the electron wave pattern. A local study of ballistic and coherent transport properties of networks of QPCs is the focus of this paper.

%

Using the metallic tip of an AFM and measuring the conductance change has already proven to be a powerful tool to locally investigate physical transport phenomena. It was successfully used to image and study the branching behaviour of electrons injected from a quantum point contact \cite{TopinkaSci,TopinkaNat,Kozikov}, fractional edge states \cite{Paradiso} and tunneling between edge channels \cite{IhnT} in the integer quantum Hall regime, localized states in graphene quantum dots \cite{Schnez} and nanoribbons \cite{Pascher, GG}, carbon nanotubes \cite{Woodside}, InAs nanowires \cite{Bleszynski}, single electron quantum dots \cite{Fallahi} and interference in quantum rings \cite{Hackens}.

In this work we study the conductance through a stadium formed by two ballistic constrictions and perturbed by an AFM tip, at small perpendicular magnetic fields. We observe fringe patterns close to the constrictions when imaging conductance versus tip position. We interpret these fringes as resulting from bottlenecks in the potential between the tip and the gates leading to quantized conductance. The fringes form a checker board pattern, which allows us to exactly determine the number of transmitted modes in the channels between the tip and the constriction edges. When the tip is scanned inside the stadium, large fluctuations of the conductance as a function of the tip position occur. Moreover, at certain tip positions the conductance is increased with respect to the conductance of the unperturbed structure. Those features cannot be interpreted in terms of the local current flow in the structure, and we have performed classical and quantum model calculations in order to compare and understand the experimental findings.

While the fluctuations inside the stadium appear already in classical transmission calculations based on electron trajectories, pointing at an origin in the chaotic dynamics of the tip-perturbed stadium, the checker board patterns result from conductance quantization. They can be qualitatively understood in terms of a network of quantized resistors and quantitatively reproduced within fully coherent model calculations using the recursive Green function method. The appearance of regions with a tip-induced increase of the conductance above the unperturbed value is consistent with results of the coherent calculation as well, and is interpreted as a signature of the coherent nature of the transport processes in our experiment.

When applying a perpendicular magnetic field we observe a transition from purely electrostatic subbands to Landau quantization while scanning across the constrictions of the stadium and thereby changing the width of the channel. We propose a model for the resulting constriction potential that is created by the combined action of the tip and the fixed gates. We fit our experimental data with this model, which is based on Ref. \cite{Beenakker1} and find for a large number of occupied subbands a good agreement between the model and the experiment.

\section{Experimental methods}

%
%
%
%
%
%

\begin{figure}[H]
\begin{center}
\includegraphics[width=11cm]{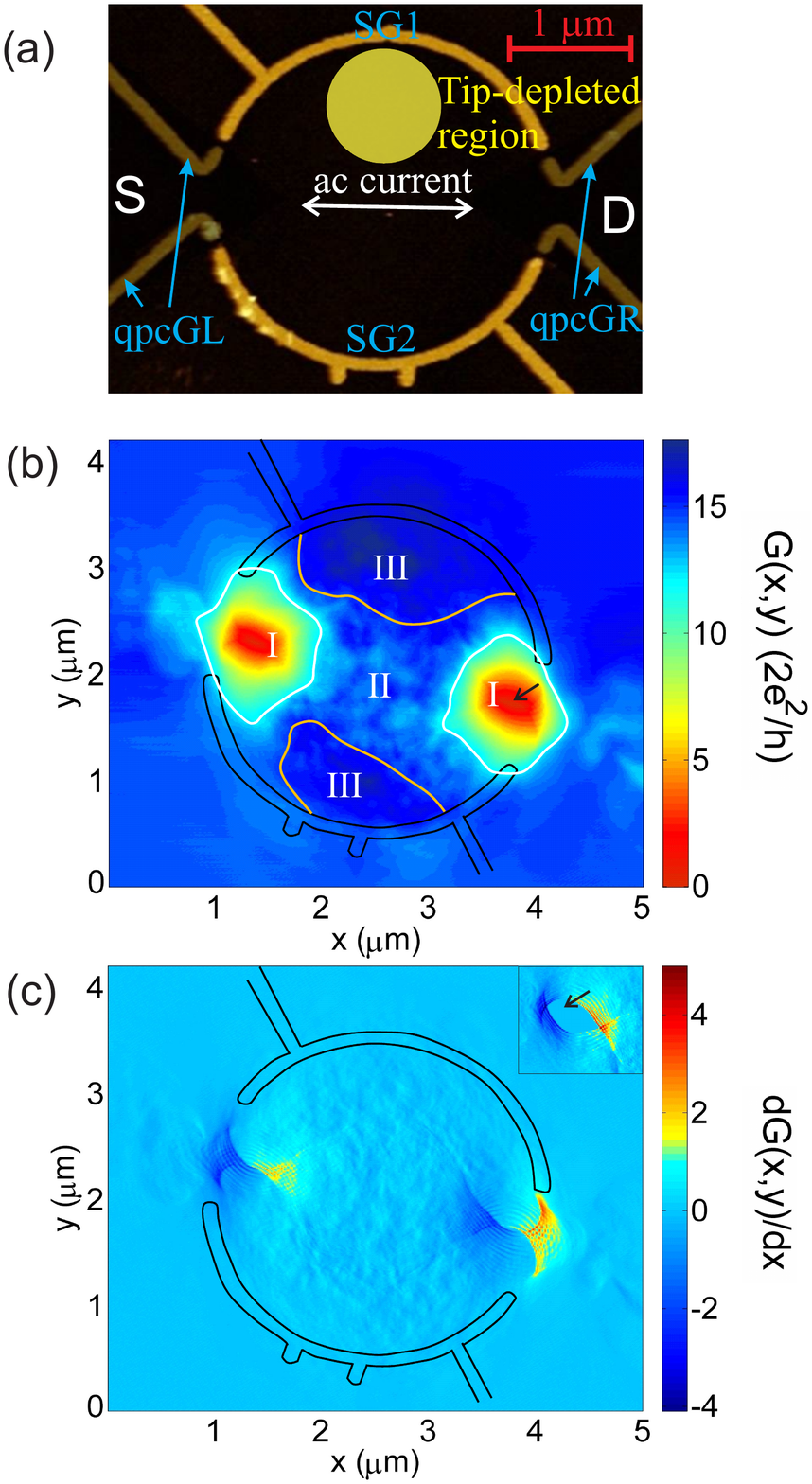}
\caption{(a) A room temperature AFM image of the sample. The stadium gates, SG1 and SG2, are biased, which is indicated by bright yellow metal. At the entrance and exit of the stadium there are two narrow QPCs (dark yellow metal) formed by the gates on the left, qpcGL, and on the right, qpcGR. The QPC top gates are grounded in this experiment. A double arrow in the center of the stadium corresponds to the ac current flowing between source (S) and drain (D). The tip-depleted disc is schematically represented by a filled yellow circle and placed as an example in the region (III) of enhanced conductance. (b) Conductance, $G$, through the stadium in units of 2e$^2$/h as a function of tip position, $(x,y)$, at zero magnetic field. The stadium is outlined by black solid lines. White and orange contour lines divide the area inside the stadium into three regions of interest, I, II and III, and correspond to the constant conductance of 12 and 15$\times$2$e^2$/h (without the tip), respectively. (c) Numerical derivative of the conductance in (a), $\rmd G(x, y)/\rmd x$, as a function of tip position. The voltage applied to the stadium gates is -1 V. Insert: The structure of the left-hand side fringe pattern when -2 V are applied to the stadium gates. The color scale is the same as in the main plot. The black arrow in the insert and in (b) indicates the lens shaped region of zero conductance.}
\label{fig:Stadium}
\end{center}
\end{figure}

Local studies of the conductance through a ballistic stadium are carried out using scanning gate microscopy (SGM). The nanostructure is fabricated on a high-mobility (800 m$^2$/Vs) GaAs/AlGaAs heterostructure. Electrons in the 2DEG located 120 nm below the surface of GaAs have an elastic mean free path of 49 $\mu$m and a Fermi energy, $E_\mathrm{F}$, of 4.4 meV. This corresponds to an electron density of $1.2\times 10^{11}$ cm$^{-2}$ and a Fermi wavelength $\lambda_\mathrm{F}=72$ nm. The stadium with a lithographic diameter of 3 $\mu$m is shown in \fref{fig:Stadium}(a) (bright yellow gates, SG1 and SG2). The entrance and exit of the stadium are constrictions, each $W=1$ $\mu$m wide, which corresponds to $N\approx27$ transmitted modes. Two narrower quantum point contacts (QPC) 300 nm wide each (shown by dark yellow gates, qpcGL and qpcGR) are fabricated in order to be able to reduce the number of transmitted modes in the stadium constrictions from 8 to 0. However, these QPC gates are not used in this work and therefore they are grounded for all the data shown in this paper.

The conductance measurements are performed in a two-terminal configuration by applying a bias of 100 $\mu$V between the source and drain contacts at 300 mK. We apply a bias of -8 V between the tip and the electron gas, which creates a local depletion region with a diameter of about 1 $\mu$m \cite{Kozikov} at a tip-surface separation of 70 nm. By scanning the tip at a constant height and simultaneously recording the conductance across the sample, 2D conductance maps, $G(x,y)$ are obtained.

\Fref{fig:Stadium}(b) shows $G(x, y)$ through the stadium when -1 V is applied to the stadium gates (the stadium is formed at a voltage smaller than -0.4 V applied to the gates). This voltage reduces the electronic width of the constrictions from 1 to 0.7 $\mu$m (see Supplementary Material) and therefore the number of transmitted modes to $N=19$. The stadium gates, SG1 and SG2, are outlined by black solid lines. One can see two distinct and symmetric regions labeled I of suppressed conductance close to the entrance and exit of the stadium. In their centers lens shaped regions are seen, in which the conductance is suppressed to zero. Here, the tip-induced potential closes the respective constriction thereby blocking electron flow through the structure. When the tip is in the center of the stadium, region II, the conductance is about 13$\times$2e$^2/$h, i.e. smaller than in the absence of the tip, $G_0\approx15 \times 2\rm e^2/$h (e.g. in the corners of the image). In region III $G$ increases to more than 16$\times$2e$^2/$h, a value that is larger than the unperturbed conductance $G_0$. The decrease of the conductance in regions I and II can be explained classically. Electrons entering the stadium from either electron reservoir are scattered by the tip back into the respective reservoir. This backscattering of electrons will reduce the conductance. The increase of $G$ in region III can not be explained in the same way. The enhanced conductance exists in this region even at low tip-sample bias voltages when the tip does not induce a depletion area. As the tip voltage is made more negative region III decreases in size and moves in the direction away from the center of the stadium.

We have performed numerical calculations of the non-interacting zero-temperature SGM-response
within a fully coherent two-dimensional tight-binding model \cite{jalabert10} using the recursive
Green function algorithm \cite{lee81,szafer89}.
With the aim to describe a situation that is as close as possible to the experiment,
we apply the method to an open ballistic circular cavity with a diameter of 2.5 $\mu$m.
The cavity is connected to wide quasi-1D
leads by openings that are similar to the quantum point contacts treated in Ref.\ \cite{jalabert10}.
The hard-wall potentials defining the QPC-like constrictions have the shape of two fingers with
circular ends of diameter 0.15 $\mu$m, facing each other with a minimal distance
of 0.8 $\mu$m. The tip induced potential is represented by a hard-wall depletion disk
with a diameter of 0.8 $\mu$m. The Fermi energy is chosen such that the Fermi wavelength
$\lambda_\mathrm{F}=73$ nm corresponds to the experimental value.
In order to keep lattice effects negligible, we have used a 2D square lattice with a lattice constant
of 5 nm, much smaller than $\lambda_\mathrm{F}$ and the length scales of the cavity.

The quantum conductance through the model cavity computed within this approach as a function of tip
position exhibits many of the experimentally observed features. Those include the lens-shaped suppression of the conductance close to the constrictions, the large conductance fluctuations as a function
of the tip position when the tip is scanned inside the cavity, and tip positions where the conductance is increased above its unperturbed value $G_0$.

Calculations of the transmission based on ballistic electron trajectories injected in the stadium under different angles in the presence of the tip show that at any tip position the conductance across the sample decreases below $G_0$. Such a behavior would also be expected in the case of incoherent diffusive transport. Interestingly, the large conductance fluctuations in region II also appear in the classical transmission, pointing at their possible relation with underlying chaotic dynamics.

The conductance increase observed in region III is reproduced only within the coherent calculations, but it does not appear in classical transport. It can thus be viewed as a signature of coherent transport through the stadium structure.

In a quantum transport picture, the behaviour of the conductance in region III can be explained as being due to a tip-induced transition from Ohmic to adiabatic transport \cite{QPCseries}. In the so-called Ohmic regime the wide cavity leads to a mixing of the modes and acts like an effective reservoir such that the total resistance is close to the series resistance of two classical resistors representing the QPCs. With the tip in regions III the width of the cavity (and thus the channel mixing) is  reduced such that the conductance of the two QPCs in series approaches that of the narrowest QPC in the adiabatic regime, becoming larger than the $G_0$ observed when the tip is outside the stadium in \fref{fig:Stadium}(b).

A significant tip-induced increase of the conductance does not occur in SGM experiments on a single QPC on a conductance plateau as the ones of Refs.\ \cite{TopinkaSci,TopinkaNat} where only negative conductance changes are reported. However, regions of positive and negative conductance change have also been observed in SGM experiments on small rings \cite{Hackens} where the unperturbed conductance is not quantized.
Though the experiments use tip voltages that represent a quite strong perturbation of the system, those observations are consistent with a recent prediction \cite{jalabert10} based on perturbation theory in the tip potential. The lowest order conductance change can be positive or negative, but its prefactor vanishes when the unperturbed structure is on a quantized conductance plateau. The second order conductance change that dominates on such a plateau is always negative. Consistently, the unperturbed conductance of our experiment is not exactly a multiple of $2e^2/h$, and regions with tip-increased conductance occur.

%

In order to emphasize the fine structure on top of these coarse conductance changes, we plot in \fref{fig:Stadium}(c) the conductance numerically differentiated with respect to the $x$-direction, $\rmd G(x, y)/\rmd x$. One can now clearly see sets of regular fringes near the constrictions of the stadium (region I) and conductance fluctuations everywhere else inside it (regions II and III). In this paper we focus on the investigation of the fringe pattern in region I. These fringes are local and roughly periodic in space: they appear when the tip is close to the constrictions formed by the top gates.

The size of the lens shaped zero conductance region depends on the electronic width of the constrictions given that the size of the tip-depleted region is fixed. The smaller the width the larger is the area of zero conductance. In the insert of \fref{fig:Stadium}(c) we show how the left-hand side fringe pattern and the lens shaped region change when -2 V are applied to the stadium gates (the fringe pattern close to the other constriction of the stadium changes in the same way). When the gate voltage is decreased from -1 to -2 V, the width of the depleted region around the gates increases, which reduces the width of the constrictions of the stadium from 700 to 600 nm. The fringes move together with the depleted region and the size of the region where the conductance is blocked increases.

\begin{figure}[t]
\begin{center}
\includegraphics[width=15cm]{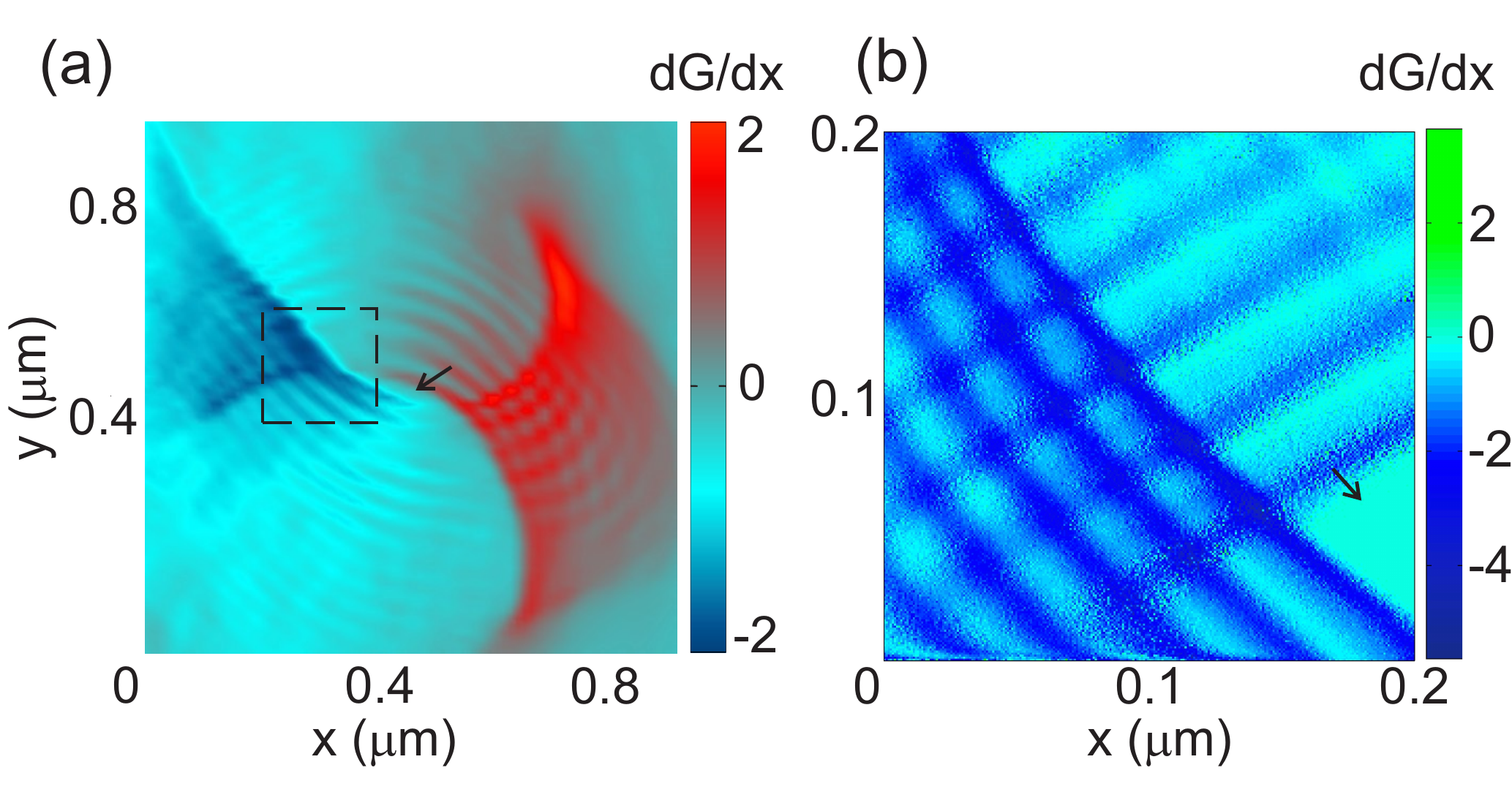}
\caption{(a) A zoom-in of a fringe pattern at the right-hand side constriction of the stadium. (b) A separate SGM measurement of the area shown in (a) by a rectangle carried out with a different AFM tip. The voltage applied to the stadium gate is -0.8 V. The arrow indicates the lens shaped region of zero conductance.}
\label{fig:FringeLens}
\end{center}
\end{figure}

\Fref{fig:FringeLens}(a) shows a zoom of the fringe pattern at the right constriction of the stadium. The fringes appear as resulting from crossing linear structures that follow the two edges of the gate fingers forming the constriction. The superposition of the two sets of lines leads to the checker board patterns to the left and to the right of the lens shaped region. \Fref{fig:FringeLens}(b) shows the left checker board pattern in detail. The triangle close to the right lower corner of the image corresponds to the zero conductance region (see black arrow; part of the lens shaped region).

\begin{figure}[t]
\begin{center}
\includegraphics[width=13cm]{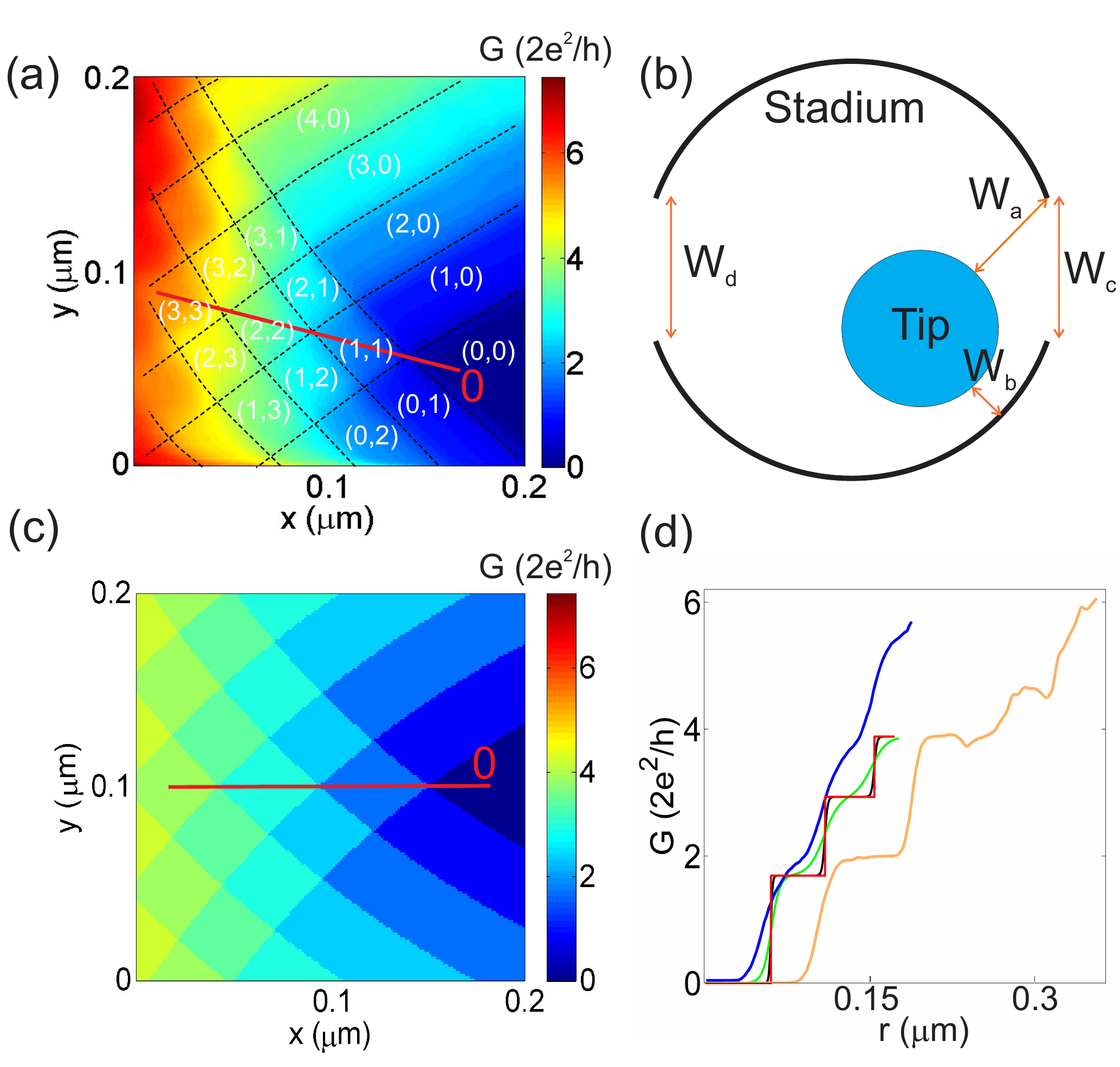}
\caption{(a) $G$ in units of 2e$^2$/h plotted as a function of tip position close to the lens shaped region (see \fref{fig:FringeLens}(b) for its numerical derivative). Dashed lines are guides to the eye. Numbers in brackets correspond to the number of transmitted modes in the constrictions formed between the tip and the stadium walls. The voltage applied to the stadium gates is -0.8 V. (b) A model used to explain the results shown in (a). (c) Results of the numerical simulations. (d) Conductance plotted as a function of distance along the red line in (a) and (c). The blue line is the experimental curve. The red curve corresponds to simulations at zero temperature and a step unit function of the transmission coefficient. The black and green curves differ from the red one by taking into account a finite temperature of $T=300$ mK and a smooth energy-dependent transmission coefficient, respectively. The number ``0'' (in red) in (a) and (c) corresponds to ``0'' in (d). The orange curve corresponds to a numerically computed conductance as a function of tip position using a fully coherent tight-binding model. The modeled curves are slightly shifted to the right for a better comparison with the experimental curve.}
\label{fig:ExpTheory}
\end{center}
\end{figure}

The absolute value of the conductance, which corresponds to the differential conductance in \fref{fig:FringeLens}(b), is shown in \fref{fig:ExpTheory}(a). In these two 2D plots one can notice that when the tip moves outside the checker board pattern, from the lens shaped region perpendicular to the fringes in the upper and lower part of the plots, $G$ changes approximately in steps of 2e$^2$/h. This change seems to occur at the dark blue stripes (fringes) in \fref{fig:FringeLens}(b) where $\rmd G/\rmd x$ is higher than that between the stripes where $\rmd G/\rmd x$ is close to zero. This behaviour is reminiscent of how the QPC conductance depends on its width: outside the checker board pattern a QPC is formed between the tip and the upper or the lower top gate of the stadium producing a set of parallel fringes below or above the lens shaped region, respectively. When the tip is inside the checker board pattern, two QPCs in parallel are formed between the tip and the top gates of the stadium (for a schematic see \fref{fig:ExpTheory}(b)).

That is why in order to explain the origin of the fringes, we consider a simple model (\fref{fig:ExpTheory}(b)), in which we describe constrictions formed at the entrance and exit of the stadium as well as those formed between the tip and the boundaries of the stadium as four incoherently coupled ballistic resistors labeled a-d. The potential of the two constrictions a and b is created by the top gate-induced potential on one side and by the tip-induced potential on the other. Electron waves bouncing off the walls of this potential self-interfere, which results in a standing wave pattern -- a wave function for each mode. The number of transmitted modes in the $i$-th constriction is estimated to be $N_i=\mathrm{int}[2W_i/\lambda_\mathrm{F}]$, where $W_i$ is the width of constriction $i$ and $i=\rm a, b, c, d$. The conductance of each of the constrictions is given by the relation $G_i=2e^2/h\times N_i$. QPCs a and b are connected in parallel, but in series with c and d giving the total conductance

\begin{eqnarray}
G_{\mathrm{Total}}=\left[\left( \frac{2e^2}{h}N_a+\frac{2e^2}{h}N_b\right)^{-1}+\left(\frac{2e^2}{h}N_c\right)^{-1}+\left(\frac{2e^2}{h}N_d\right)^{-1}\right]^{-1}.
\label{eqn:ModelResistance}
\end{eqnarray}

When the tip moves, only the number of modes $N_\mathrm{a}$ and $N_\mathrm{b}$ changes in our model. Beyond that the model assumes that the temperature is zero and that the transmission for each mode is a step function of energy, i.e. $N_i$ is an integer. The dependence of $G_{\mathrm{Total}}$ on tip position resulting from the model is plotted in \fref{fig:ExpTheory}(c). The value of the Fermi wavelength was determined from classical Hall effect measurements, i.e. from the 2DEG density, and the widths of the constrictions $W_\mathrm{c}$ and $W_\mathrm{d}$ are both taken to be equal to 0.8 $\mu$m (for the stadium gate voltage of -0.8 V the size of the constriction is about 0.8 $\mu$m).

\begin{figure}[h]
\begin{center}
\includegraphics[width=13cm]{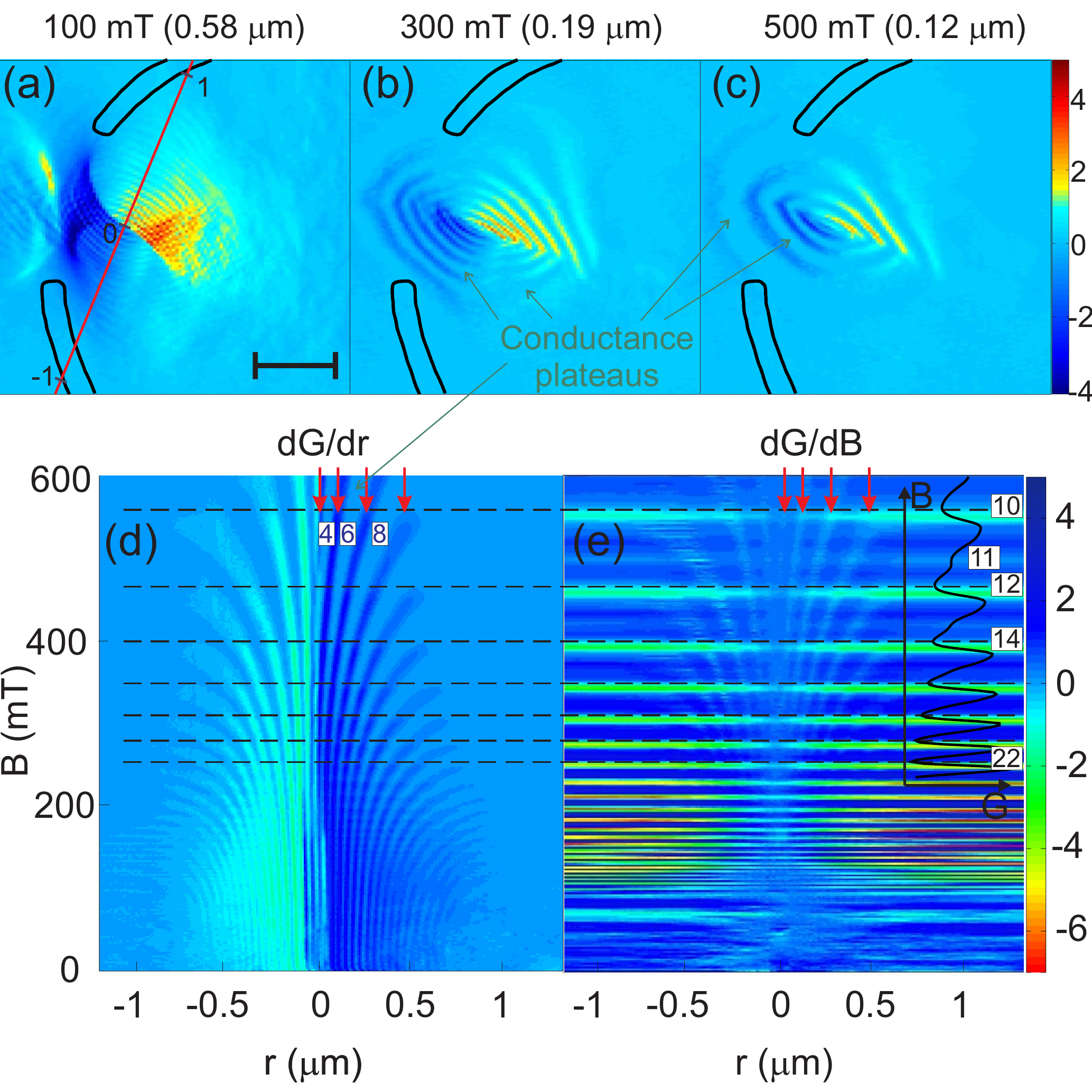}
\caption{(a)-(c) Evolution of the left-hand side fringe pattern with a perpendicular magnetic field. The derivative of the conductance, $\rmd G(x, y)/\rmd x$, is plotted as a function of tip position, $(x, y)$, at 100, 300 and 500 mT. The stadium walls are indicated by the black solid lines. The scale bar corresponds to 500 nm. Numbers in brackets correspond to values of the classical cyclotron radius. Numbers -1, 0 and 1 along the red line in (a) correspond to the distance in microns. They are the same as on the horizontal axis in (d). (d) The conductance along the red line in (a) plotted as $\rmd G(B, r)/\rmd r$ as a function of magnetic field, $B$, and distance, $r$. (e) The conductance along the red line in (a) plotted as $\rmd G(B, r)/\rmd B$. The black solid curve shows the conductance, $G$ along a short black horizontal axis, (Shubnikov-de Haas) oscillations as a function of magnetic field (black vertical axis) in the range between 230 and 600 mT. The horizontal dashed line indicates positions of conductance minima in this curve and the filling factors in the bulk. A smooth background was subtracted from the black curve. Red arrows in (d) and (e) indicate transitions between conductance plateaus. Numbers in squares in (d) and (e) correspond to filling factors in the constriction and in the bulk, respectively. In (a)-(e) the voltage applied to the stadium is -1 V.}
\label{fig:G(B,x)}
\end{center}
\end{figure}

The model describes the experimental results quite well on a qualitative level, which is seen by comparing the calculated checker board pattern in \fref{fig:ExpTheory}(c) with the measured one in \fref{fig:ExpTheory}(a). To check the quantitative agreement between the two images, we plot in \fref{fig:ExpTheory}(d) the measured (blue) and calculated (red) conductance along the red line shown in (a) and (c). In addition to these two curves, a finite temperature of 300 mK (black curve) and a smooth energy-dependent transmission coefficient (green curve) are also taken into account in the calculation for comparison. The two additional curves were calculated using a saddle-point model based on the experimentally determined subband spacings for qpcGR \cite{Halperin, Buttiker, Rossler} (see \fref{fig:Stadium}(a)). We then assumed that all constrictions, a-d, behave in the same way in the structure under investigation. One can see that the absence of clear plateaus in the experimental curve is due to the smooth energy-dependent transmission and to the finite temperature of 300 mK, but the latter is dominated by the first. The slopes of the experimental and simulated curves are not exactly the same due to differences in the potentials of the constrictions.
In order to eliminate the difference in the plateau conductance, the number of transmitted modes in the stadium constrictions, $N_c$ and $N_d$, would have to be taken infinite (see Supplementary Material). It is also important to note that in the model the lens shaped regions are in the centers of the stadium constrictions. In the experiment they appear inside the stadium, which is probably due to screening effects. Although the model agrees qualitatively with the experiment, it should be used with care given the assumptions made. The real potential is not a hard-wall potential as we assumed, which means that all four resistances change with tip position. The use of only four QPCs is a rough approximation. However, the model does qualitatively explain our observations. Thus, the fringes at the stadium constrictions seen in $\rmd G/\rmd x$ are basically the result of conductance quantization plateaus in $G(x, y)$ arising from a network of quantum point contacts.

The fully coherent calculation using the recursive Green function method reproduces the checker board-like patterns in \fref{fig:FringeLens} and \ref{fig:ExpTheory}.
In \fref{fig:ExpTheory}(d), we show the numerical results (orange line) for tip positions along a straight line similar to the red line in \fref{fig:ExpTheory}(d)) that is
parallel to the electron propagation.
The qualitative agreement with the experimental blue curve is striking.
However, there is a difference in the values of the conductance of the third plateau. The reason can be that the tip being inside of the stadium makes the dynamics in the stadium chaotic and thereby affects the total conductance not only by opening or closing the constriction as it is in the simple model described previously (\fref{fig:ExpTheory}(a)). This makes the system extremely sensitive to details of the geometry and other parameters, in particular in the fully coherent calculations where the temperature was taken to be zero. While the observed conductance step heights are well reproduced by the theory, the tip positions where subbands are opened when moving the tip away from the center of the constrictions show small quantitative deviations due to the hard wall potentials used to model the tip depletion
disk and the gate-defined constrictions. Moreover, the capacitive coupling between the tip and
the top-gates of the sample, not taken into account in the simulation, might have an influence in the experiment.

The stripes in \fref{fig:FringeLens}(b) or \ref{fig:ExpTheory}(c) and diamonds formed by the crossing stripes show how many modes are open in the constrictions formed between the tip and the stadium walls. For example, following the red line in \fref{fig:ExpTheory}(a) or (c) we start from the zero conductance region. Then the tip moves to the first diamond-like area. This corresponds to the situation when the two constrictions ($a$ and $b$ in \fref{fig:ExpTheory}(b)) have one open mode each, $(N_\mathrm{a}, N_\mathrm{b})=(1, 1)$, and the conductance is constant inside this area in the ideal case shown in \fref{fig:ExpTheory}(c). For the second diamond two QPC modes are open in both constrictions, (2, 2), and so on. Thus, when moving along the red line the tip passes sequentially through the (0, 0) to the (3, 3) regime. In other areas of the image $N_\mathrm{a}\neq N_\mathrm{b}$. For example, outside the checker board pattern one of the two numbers of modes, $N_\mathrm{a}$ or $N_\mathrm{b}$, is equal to zero, whereas the other one increases starting from 0 (the lens shaped region). The rectangular form of these regions in the experiment is due to slightly different lever arms of characteristic induced potentials of the stadium top gates.



\Fref{fig:G(B,x)}(a)-(c) shows the evolution of the fringe patterns in $\rmd G/\rmd x$ (only one is shown as an example) in a perpendicular magnetic field. As the magnetic field increases, the number of observable conductance plateaus decreases and their width increases.
To see precisely the effect of the magnetic field on the fringe patterns with higher resolution in $B$-field, we plot in \fref{fig:G(B,x)}(d) the derivative of the conductance, $\rmd G(B, r)/\rmd r$, as a function of magnetic field and distance along the red line shown in \fref{fig:G(B,x)}(a). At zero magnetic field a set of fringes, which correspond to the transition between conductance plateaus, appear close to the constrictions like in \fref{fig:Stadium}(c). As the magnetic field increases from 0 to 600 mT the plateaus' width increases. Above 100 mT the number of observable plateaus decreases. The effect of the magnetic field on the number of occupied subbands becomes important when the classical cyclotron orbits do not fit into the width of the constriction, $W_i$. When the tip is outside the constriction (along the red line in \fref{fig:G(B,x)}(a)) and the cyclotron radius, $r_\mathrm{c}=\hbar k_\mathrm{F}/(eB)$, is bigger than $W_c$ or $W_d$, then the geometric confinement dominates over the magnetic confinement and the number of conductance plateaus is independent of $B$. The condition $r_\mathrm{c}=W$ in our experiment is fulfilled at about 100 mT where $r_\mathrm{c}\approx 600$ nm, and $W\approx 700$ nm when -1 V is applied to the stadium gates. When the tip approaches the constriction, $W_c$ or $W_d$ further decreases becoming equal or even smaller than $r_\mathrm{c}$. This means that below 100 mT the number of the plateaus should not change, which is in agreement with our experiment, \fref{fig:G(B,x)}(d). At higher magnetic fields, when $r_\mathrm{c}<W/2$ (fulfilled already at 200 mT), the number of plateaus decreases as a function of $r$ due to the magnetic depopulation of subbands as also seen in the figure. Around $\pm$ 0.7 $\mu$m at low magnetic fields the fringes disappear completely, because the tip is outside the constriction. The estimated corresponding distance from the center of the constriction is $\mid W_c$/2+$R_{\mathrm{tip}}\mid=0.35+0.5=0.85$ $\mu$m, which is close to 0.7 $\mu$m. The difference may originate from the radius of the tip-depleted region, for which the value of 0.5 $\mu$m is an estimation.

\begin{figure}[t]
\begin{center}
\includegraphics[width=13cm]{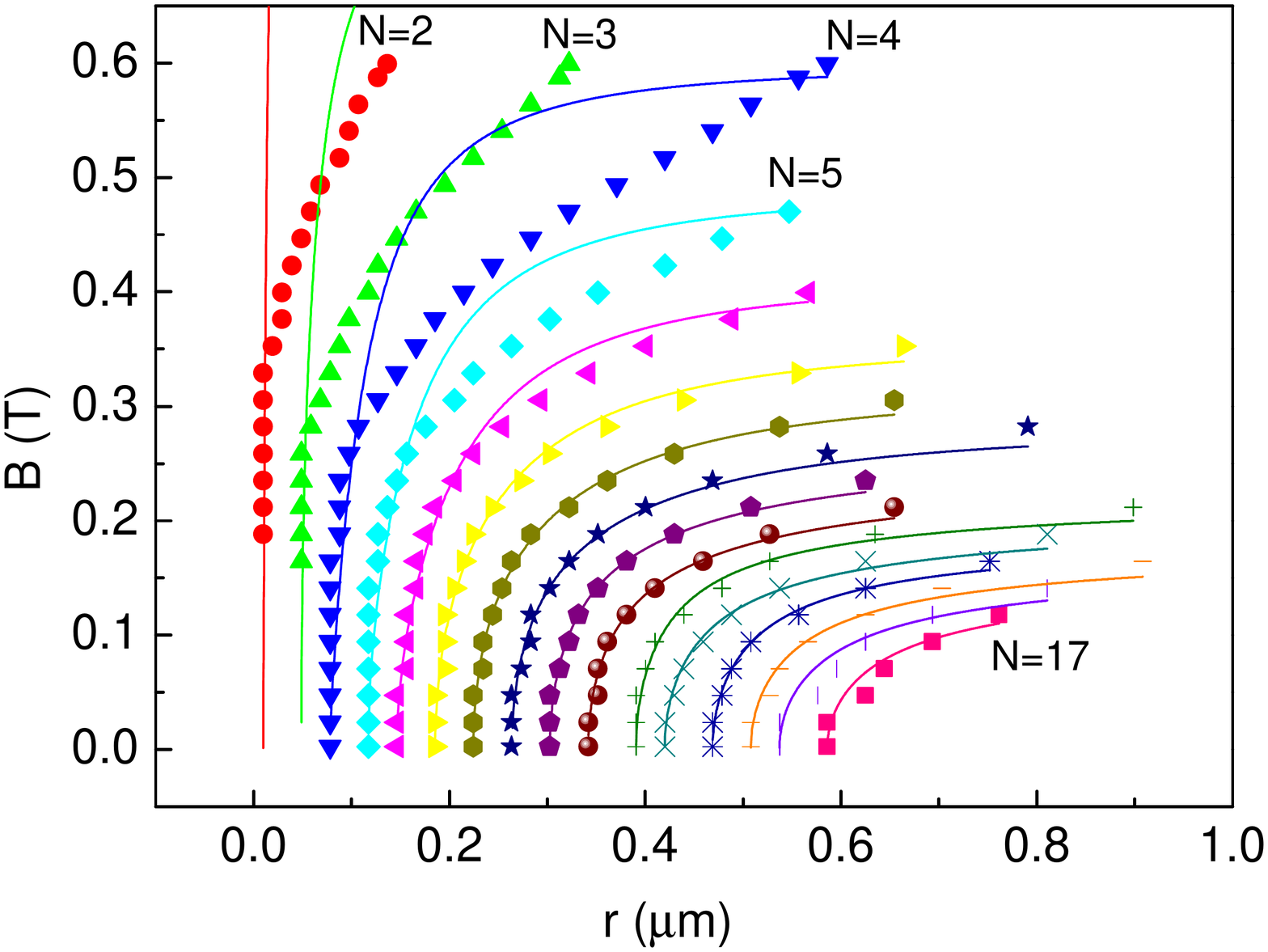}
\caption{Fitting of the experimental data (points) extracted from \fref{fig:G(B,x)}(d) to the model (solid lines) described in the main text.}
\label{fig:BfanFit}
\end{center}
\end{figure}

\Fref{fig:G(B,x)}(e) shows the derivative $\rmd G(B, r)/\rmd B$ as a function of distance and magnetic field along the same red line in (a). In comparison with (d), horizontal lines are now clearly seen. They correspond to regions of the magnetic field between extrema in the bulk Shubnikov-de Haas oscillations  (an example is illustrated by the black solid line when the tip is outside the constriction). Bulk filling factors (numbers in squares) are assigned to the conductance minima. The conductance along any horizontal cut in \fref{fig:G(B,x)}(e) at a constant finite magnetic field exhibits clear plateaus (see \fref{fig:G(B,x)}(a)-(c)) each of which corresponds to a certain filling factor in the constriction. The boundaries of these plateaus are marked by the red arrows for a particular magnetic field as an example. For instance, for the uppermost horizontal dashed line, when the tip is outside the constriction the filling factor is that of the bulk, i.e. 10. As the tip moves closer to the center of the constriction the edge channels are brought closer to each other and are closed one by one, which is seen as a stepwise reduction of the filling factor from 10 to 4 (see \fref{fig:G(B,x)}(d)). Stripes between the fringes widening with magnetic field (bent green or dark blue) in \fref{fig:G(B,x)}(d) correspond to one particular filling factor (subband occupation). The stripes are vertical at low magnetic fields and bent towards larger $r$ at high $B$-fields (see the model below).

\Fref{fig:BfanFit} shows experimental points for the maxima of the fringe pattern extracted from \fref{fig:G(B,x)}(d). Each set of points of the same color mark the transitions between integer number of subbands starting from $N=2$ (the closest to $r=0$ $\mu$m) to $N=17$. We propose a model to explain the depopulation of the subbands as a function of magnetic field and tip position, which is an extension of the model of Beenakker \cite{Beenakker1}. The constriction is formed by the superposition of the gate-induced and the tip-induced potential. The gate-induced potential is fixed in space and that formed by the tip can be moved. The size of the constriction is changed by moving the tip above the surface. In our model we approximate the potential of the gate and the tip along the narrowest section of the constriction in confinement direction $x$ by a $1/x^2$-power law as
\begin{eqnarray}
V(x)=\left[\frac{1}{x^2}+\frac{V_0}{(x-r)^2}\right]U,
\label{eqn:BfanModelPotential}
\end{eqnarray}
where $r$ is the tip position, $V_0\propto V_\mathrm{tip}/V_g$, $V_\mathrm{tip}$ is the tip bias, $V_g$ is the gate voltage, $U\propto V_g$ in units of eVm$^2$. By expanding this potential around its minimum, $x_\mathrm{min}$, one obtains in lowest order the resulting parabolic transverse potential
\begin{eqnarray}
V(x)=V_\mathrm{min}+\frac{1}{2}m\omega_0^2 (x-x_\mathrm{min})^2=\frac{U}{\alpha x_\mathrm{min}^2}+\frac{1}{2}\frac{6U}{(1-\alpha)x_\mathrm{min}^4} (x-x_\mathrm{min})^2,
\label{eqn:BfanModelExpandedPotential}
\end{eqnarray}
where $V_\mathrm{min}(r)$ is the minimum of $V(x)$ at $x_\mathrm{min}$, $\alpha=x_\mathrm{min}/r$ describes a relative motion of the potential minimum with respect to the tip motion and $\hbar \omega_0(r)$ is the subband spacing at $B=0$. We can see that this model incorporates the fact that the potential minimum in the constriction between the tip and the gate is lifted as the tip approaches the gate. At the same time, the position of the minimum shifts proportionally to the tip position with a lever arm $\alpha$. Furthermore, the parabolic confinement $\omega_0$ becomes sharper as the tip approaches the gate. These three effects are incorporated by the two parameters $\alpha$ and $U$.

The model takes into account the tip-position-dependent potential minimum in space, $x_\mathrm{min}$, and in energy, $V_\mathrm{min}(r)$. The minima of the magnetoelectric subbands have energies $E_N=(N+1/2)\hbar \omega(r,B) + V_\mathrm{min}(r)$, where $\omega(r,B)^2=\omega_0(r)^2+\omega_c(B)^2$, $\omega_c$ is the cyclotron frequency and $N>0$ is an integer labeling the subbands \cite{Beenakker1}. The $N$\textsuperscript{th} magnetoelectric subband is depopulated at tip position $r_N$ and magnetic field $B_N$ if $E_N(r_N, B_N)=E_F$. From this expression the dependence of the magnetic depopulation field $B_N$ on tip position $r_N$ is

\begin{eqnarray}
B_N = \frac{nh}{2e}\sqrt{\left\{ -\left(\frac{K}{r_N^4}\right)^2  +   \left[  \frac{1-   \left(\frac{\tilde{r}}{r_N}\right)^2  +   \frac{(N+0.5)K}{(\tilde{r}^2 r_N^2)^2}  }{N+0.5}    \right]^2                   \right\}},
\label{eqn:BfanModelB}
\end{eqnarray}
where $n$ is the carrier density in the absence of the tip, $\tilde{r}$ is the tip position at $B=0$ at which $N$ is an integer for different $N$, $K$ is a fitting parameter.

The results of the fitting procedure using equation (\ref{eqn:BfanModelB}) with the same value of $K$ for all curves are shown in \fref{fig:BfanFit} by solid lines. When seven and more subbands are occupied the agreement between model and experiment is very good. For lower subband occupation the modelled curve starts deviating from the experimental points. In our model this deviation becomes stronger as $N$ decreases. A more elaborate model would include carrier-carrier interactions and electron interference.

%
%




\section{Conclusion}

We have presented experimental results of local studies of the conductance through a ballistic cavity connected to the leads by two constrictions using scanning gate microscopy in the presence of small perpendicular magnetic fields. We observed strong tip-position dependent fluctuations of the conductance when the tip was scanned inside the cavity, regions where the conductance was increased above the unperturbed conductance of the structure, and checker board shaped fringe patterns when the tip was close to one of the constrictions. These patterns allowed to precisely control the number of transmitted modes in the channels formed between the tip potential and the confinement caused by the top gates, which may in the future be used to image Aharonov-Bohm oscillations in the quantum Hall regime. Investigating the fringe patterns in a perpendicular magnetic field we observed a transition from the electrostatic to magnetic depopulation of subbands. This effect was seen in spatially resolved images where narrow fringes were present at low magnetic fields, which gradually transformed into wide conductance plateaus at higher $B$-fields. The plateaus corresponded to integer filling factors. Classical and quantum simulations described very well most of our observations.

The observed features in the tip-induced conductance changes and the comparison with the model calculations clearly indicate that our device was operated in the coherent quantum transport regime. Moreover, our modelling allowed to understand in detail the origin of the observed conductance changes. Most of the local features seen in our experimental conductance images cannot directly be interpreted in terms of a corresponding local current flow as it was suggested \cite{TopinkaNat,TopinkaSci} in the context of single QPCs operated on a quantized conductance plateau. Though we cannot measure the local current density in the absence of the tip, it seems obvious that the checker board fringe patterns observed close to the QPCs and the tip-induced enhancement of the conductance above the unperturbed value do not reflect the behaviour of local current flow in the unperturbed device. Since our data do not show a well-defined conductance plateau, the absence of an interpretation of the conductance change in terms of local current flow is consistent with a recent theoretical study \cite{gorini13}, perturbative in the tip potential, where the conductance change could be unambiguously related to the local current density only in the case of a structure having spatial symmetry and being operated on a quantized conductance plateau.

Quantitative deviations between the experimental results and model calculations could be due to interaction effects which were not taken into account in the models. Further experiments and refined modelling are needed to clarify this.

\section{Acknowledgements}

We are grateful for fruitful discussions with Markus B\"{u}ttiker, Rodolfo Jalabert, Cosimo Gorini and Bernd Rosenow. We acknowledge financial support from the Swiss National Science Foundation and NCCR ``Quantum Science and Technology" and the French National Research Agency ANR within project ANR-08-BLAN-0030-02.

\section*{Appendix}

\appendix
\setcounter{section}{1}

\begin{figure}[t]
\begin{center}
\includegraphics[width=15cm]{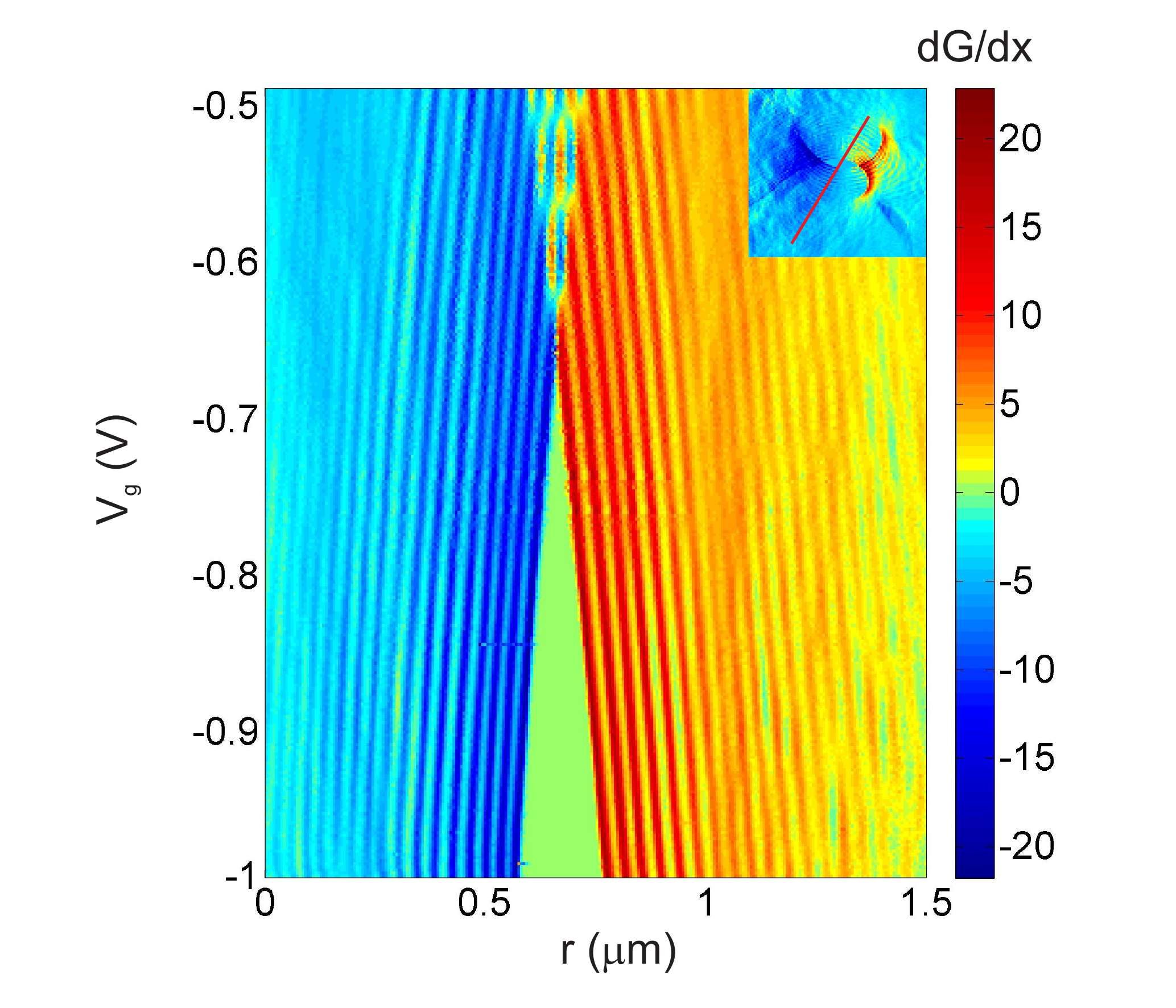}
\caption{Derivative of the conductance with respect to distance as a function of distance and top gate voltage. The insert shows the right-hand side fringe pattern. Conductance in the main graph was measured along the red line.}
\label{fig:GateWidth}
\end{center}
\end{figure}

In order to determine the width of the stadium constriction at different top gate voltages, the conductance was measured as a function of voltage applied to the stadium gates along the red line shown in the insert to \fref{fig:GateWidth}. The green triangle in the main graph at $x\approx0.6$ $\mu$m is the zero conductance region (in real space it is a lens shaped region). Tilted lines on both sides of this region correspond to sets of fringes also observed in \fref{fig:Stadium}(c). As the gate voltage changes, the fringes move together with the depleted region around the top gates. They move linearly with distance due to the constant slope of the lines. Therefore, the slope of the lines allows determining how the depleted region around the top gates changes with gate voltage. One can see from the main figure that the size of the constriction changes with the gate voltage as $\Delta r/\mu \mathrm{m}=0.25\Delta V_\mathrm{g}/\mathrm{V}$.

The voltage at which the stadium constrictions are formed is -0.4 V. In this case their widths are roughly equal to the lithographic width, which is 1 $\mu$m each. In the presented measurements voltages of -0.8 and -1.0 V were applied to the top gates, which correspond to the size of the constrictions of $1-2\cdot0.25\cdot(0.8-0.4)=0.8$ $\mu$m and $1-2\cdot0.25\cdot(1.0-0.4)=0.7$ $\mu$m, respectively.

The lever arms of the two top gates are slightly different from each other, seen as different slopes of the lines to the left and to the right of the zero conductance region. For the lower gate $\Delta x_\mathrm{L}/\mu \mathrm{m}=0.25\Delta V_\mathrm{g}/\mathrm{V}$ and for the upper gate $\Delta x_\mathrm{U}/\mu \mathrm{m}=0.3\Delta V_\mathrm{g}/\mathrm{V}$. However, the sizes of the stadium constrictions determined with the different and the same lever arms differ only by 2$\%$ for the top gate voltage of -0.8 V and by 4$\%$ for -1.0 V. Such errors will not change the results presented in this work. We therefore consider the lever arms of the two top gates to be the same in our paper. Changing the orientation of the red line to vertical results in the same lever arms.

\begin{figure}[t]
\begin{center}
\includegraphics[width=15cm]{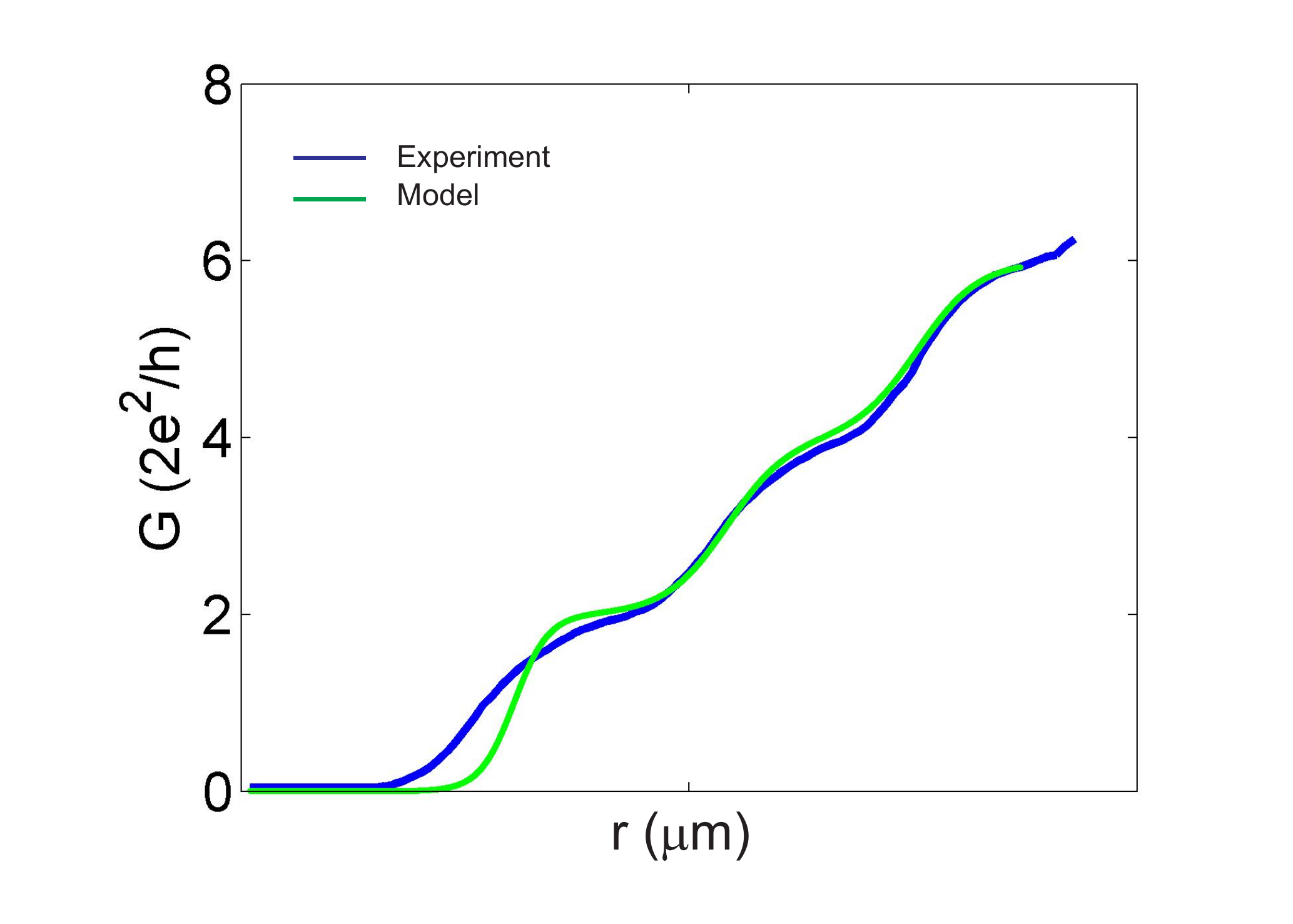}
\caption{Conductance plotted as a function of distance along the red line in \fref{fig:ExpTheory}(a) and (c). The blue line is the experimental curve. The green curve corresponds to the model. A smooth energy-dependent transmission coefficient is taken into account. In the model the resistance of the stadium constrictions $R_a$ and $R_b$ are neglected.}
\label{fig:ExpTheoryEtra}
\end{center}
\end{figure}

\Fref{fig:ExpTheoryEtra} shows a comparison between the measured conductance trace along a line perpendicular to the fringe pattern and the constriction model similar to the one in \fref{fig:ExpTheory}(d). This time only the constrictions formed between the tip and the stadium are taken into account. The contact resistance subtracted to obtain the experimental curve was increased by 20$\%$ compared to the model discussed in the main text in order to align the third conductance plateau in the two curves in the figure. This added contact resistance accounts for the small contribution to the total resistance coming from the two stadium constrictions. The difference between the curves around the first plateau arises (as mentioned in the main text) from the saddle point potential used in the model.


\end{document}